\documentclass[twocolumn,twocolappendix]{openjournal}
\usepackage{float}
\usepackage{xcolor,bm}
\usepackage{amsmath}
\usepackage{placeins}
\usepackage{savesym}
\savesymbol{tablenum}
\usepackage{siunitx}
\restoresymbol{SIX}{tablenum}
\DeclareSIUnit\angstrom{\text {Å}}

\begin{document} 

\title{Rapid identification of lensed type Ia supernovae with color-magnitude selection \vspace{-0.5in}}
\shorttitle{Rapid identification of lensed SNe Ia}
\shortauthors{Mane et al.}

\author{Prajakta Mane$^1$, Anupreeta More$^{2,3}$, Surhud More$^{2,3}$}
\affiliation{$^{1}$Department of Physical Sciences, Indian Institute of Science Education and Research, Mohali, India}
\affiliation{$^{2}$The Inter-University Centre for Astronomy and Astrophysics, Post bag 4, Ganeshkhind, Pune 411007, India}
\affiliation{$^{3}$Kavli Institute for the Physics and Mathematics of the Universe (WPI), 5-1-5 Kashiwanoha, Kashiwa-shi, Chiba 277-8583, Japan}

\begin{abstract}
Strongly lensed type Ia supernovae (SNe Ia) provide a unique cosmological probe to address the Hubble tension problem in cosmology. In addition to the sensitivity of the time delays to the value of the Hubble constant, the transient and standard candle nature of SNe Ia also enable valuable joint constraints on the model of the lens and the cosmological parameters. The upcoming Legacy Survey of Space and Time (LSST) with the Vera C. Rubin Observatory is expected to increase the number of observed SNe Ia by an order of magnitude in ten years of its lifetime. However, finding such systems in the LSST data is a challenge. In this work, we revisit the color-magnitude (CM) diagram used previously as a means to identify lensed SNe Ia and extend the work further as follows. We simulate LSST-like photometric data ($rizy$~bands) of lensed SNe Ia and analyze it in the CM parameter space. We find that a subset of lensed SNe Ia are redder compared to unlensed SNe Ia at a given magnitude, both in the rising and falling phases of their light curves and for SNe up to $z=3$. We propose a modified selection criterion based on these new results. We show that the contamination coming from unlensed core-collapse (CC) SNe is negligible, whereas a small fraction of lensed CC SNe types Ib and Ic may get selected by this criterion as potential lensed SNe. Finally, we demonstrate that our criterion works well on a wide sample of observed unlensed SNe Ia, a handful of known multiply-imaged lensed SNe systems, and a representative sample of observed CC SNe as well as super-luminous supernovae.
\end{abstract}
\maketitle

\section{Introduction}
Gravitational lensing, deflection in the path of light from a source in the presence of deep gravitational potential, can lead to the formation of multiple images of the same source. The multiple images appear in the sky separated by a time delay. The time delays between multiple images in strong lensing yield a direct measure of the cosmological distances involved in the lens system and thus constrain cosmological parameters such as the Hubble constant \citep[$H_0$;][]{refsdal1964possibility}. This method of constraining $H_0$ - time-delay cosmography - has so far primarily been accomplished with lensed quasars, \citep[e.g.,][]{suyu2017h0licow, wong2017h0licow, chen2019sharp, birrer2020tdcosmo, shajib2020strides, wong2020h0licow}. \cite{wong2020h0licow} inferred $H_0$ using 6 lensed quasar systems and gave a joint constraint of 73.3$^{+1.7}_{-1.8}$ km s$^{-1}$ Mpc$^{-1}$. However, there are challenges involved in time-delay cosmography with lensed quasars. The light curves of quasars are stochastic and heterogeneous, thus typically requiring years of monitoring of lensed quasar systems to calculate precise time delays \citep[][]{liao2015strong}. Since quasars typically outshine the host and the lens galaxy light, reconstructing the lensing potential becomes challenging.

In contrast, lensed supernovae (SNe) can be more useful for time-delay cosmography compared to quasars due to their transient nature and better-constrained light curves. SNe light curves are relatively well-studied and predictable, which makes it easier to constrain their time delays with shorter observing campaigns \citep[][]{goldstein2018precise}.  
The transient nature of SNe enables better follow-up imaging of the host and lens galaxy without contamination from the SN light as it eventually fades away. These measurements can help constrain the lens mass model.
Additionally, type Ia supernovae (SNe Ia) have remarkably homogeneous light curves among all SNe types with almost identical luminosity within their class. This standardizable nature of SNe Ia enables constraints on the absolute magnification factors of individual images, which can be valuable in minimizing the mass-sheet degeneracy and resulting in improved $H_0$ measurements \citep[][]{oguri2003gravitational}. Alternatively, it also allows joint constraints on the lens model and the cosmological parameters \citep[e.g.,][]{oguri2003gravitational, linder2004strong, linder2011lensing}. For further discussion on the suitability of lensed SNe for time-delay cosmography and various methods to identify such systems in large-scale surveys, see recent reviews \cite[e.g.,][]{oguri2019strong, liao2022strongly, suyu2024strong}.

While lensed SNe show great potential for time-delay cosmography due to their aforementioned properties, such events are very challenging to discover and follow up before they fade away. It is also particularly difficult to distinguish them from other transients and choose for further follow up. Currently, our sample of known strongly lensed multiply imaged SNe contains only eight systems: PS1-10afx \citep[][]{quimby2014detection}, SN Refsdal \citep[][]{kelly2015multiple}, iPTF2016geu \citep[][]{goobar2017iptf16geu}, SN Requiem \citep[][]{rodney2021gravitationally}, C22 \citep[][]{chen2022shock}, SN Zwicky \citep[][]{goobar2023uncovering}, SN H0pe \citep[][]{frye2024jwst}, and SN Encore \citep[][]{pierel2024lensed}. Out of the above, PS1-10afx, SN Zwicky, SN Requiem, iPTF2016geu, SN H0pe, and SN Encore either show strong evidence of type Ia nature \citep[][]{quimby2014detection} or are spectroscopically confirmed to be of type Ia \citep[][]{goobar2017iptf16geu, rodney2021gravitationally, goobar2023uncovering, frye2024jwst, dhawan2024spectroscopic}. Inferring $H_0$ from these systems has been primarily limited owing to insufficient follow-up data, too short time delays between multiple images, and systematics in lens mass modeling. Until now, SN Refsdal was able to provide $H_0$ constraints at 66.6$^{+4.1}_{-3.3}$ km s$^{-1}$ Mpc$^{-1}$ \citep[][]{kelly2023constraints} and SN H0pe provided constraints at 75.4$^{+8.1}_{-5.5}$ km s$^{-1}$ Mpc$^{-1}$ \citep[][]{pascale2025sn}. 

Legacy Survey of Space and Time (LSST) to be conducted at the Rubin Observatory \citep[][]{ivezic2011large, ivezic2019lsst} will be a powerful SN factory, discovering $\mathcal{O}(10^6)$ of SNe Ia in 10 years of its survey life \citep[][]{abell2009lsst, ivezic2019lsst}. It is also expected to find several hundreds of lensed SNe Ia in its lifetime \citep[e.g.,][]{oguri2010gravitationally, quimby2014detection, goldstein2016find, wojtak2019magnified, arendse2024detecting}. \cite{arendse2024detecting} predict that lensed SNe Ia detected within about three years of LSST operations will lead to a 1.5\% precision in the measurement of $H_0$. The prospect of precise, independent measurement of $H_0$ is particularly important given the Hubble Tension - the $4~\sigma-6~\sigma$ disagreement in the value of $H_0$ as measured using early-time and late-time probes \citep[e.g.,][]{freedman2001final, aghanim2020planck, wong2020h0licow, riess2021cosmic}. Due to the cosmological importance and rarity of occurrence of lensed SNe Ia, it is paramount to devise effective methods to identify such systems from the LSST data. 

\subsection{Color-magnitude selection criterion}\label{ssec:cmaintro}
As the observed lensed SNe population will primarily come from higher redshifts compared to unlensed SNe, the peak of the spectral energy distribution for SNe Ia will shift out of the observer frame bluer bands, giving redder observed colors to lensed SNe Ia. Additionally, for lensed and unlensed SNe Ia at the same redshifts, lensed SNe Ia will highly likely appear brighter because of the lensing magnification. Owing to this, \cite{quimby2014detection} proposed the idea of exploring the color-magnitude (CM) space of SNe to identify promising strongly lensed SNe candidates. They proposed a ``red limit'', a limit on the color of unlensed SNe for a given apparent magnitude \citep[the bold black curve in figure 4 of][]{quimby2014detection}, as a promising criterion to select only lensed SNe. This criterion selects the SNe that lie above the red limit as potential lensed SNe Ia. The original red limit was proposed to select unresolved lensed SNe on the rising edges of their light curves. 

More recently, \cite{magee2023search} utilized color-magnitude selection as one of the methods to identify lensed SNe from the first four years of transient detections by the Zwicky Transient Facility (ZTF). \cite{townsend2025candidate} extended this work to run lensed SNe search methods on a more extensive ZTF trasient dataset reaching fainter magnitudes and applied similar color-magnitude cuts as selection criteria for potential lensed SNe. Expanding to future surveys, \cite{arendse2024detecting} explored the use of color-magnitude selection within LSST-like SN Ia simulations, evaluating its effectiveness in separating the populations of lensed and unlensed SN Ia in the upcoming LSST data. These studies highlight the potential of color-magnitude selection in lensed SN searches under different cases (selection cuts for ZTF/LSST survey properties, how microlensing affects its efficiency, the effect of non-Ia SNe), while also identifying its limitations.

In the current work, we extend the original study by \cite{quimby2014detection} and systematically investigate photometric data of unlensed and lensed SNe Ia and core-collapse (CC) SNe in the CM parameter space. We propose red limits for our sample that select lensed SNe Ia candidates, both unresolved and resolved, on both the rising and falling phases of the light curve. Since the focus of this study is to identify lensed SNe type Ia, we consider other types of (un-)lensed SNe as contaminants and investigate the contamination from a population of simulated CC SNe and find the subtypes that act as primary contaminants. Finally, To verify the applicability of the proposed criterion on the observed SNe data, we find real unlensed SNe Ia and CC SNe from a variety of archival surveys and show that the proposed selection criterion successfully eliminates a majority of this population. Along with this, we also check for contamination by a representative sample of observed superluminous supernovae (SLSNe). 

We outline our (un-)lensed SNe Ia and CC SNe simulation procedure in Sec. \ref{sec:SNe_sim} and describe the details of the observed SNe Ia, CC SNe, and SLSNe data used from various published surveys in Sec. \ref{ssec:observed_SNeIa}. We present the results from the investigation of the CM diagram for various SNe samples in Sec. \ref{sec:results}, and describe the conclusions along with discussion in Sec. \ref{sec:conclusion}. We use AB magnitudes throughout this work.

\section{Simulating (un-)lensed supernovae}\label{sec:SNe_sim}
In this section, we describe the assumptions and the details of the pipeline employed to generate the strong lensing observables - time delay, image positions, and magnifications - for individual images and the details of the (un-)lensed light curve simulations. For strong lensing simulations, we draw the lens galaxies from the HSC galaxy catalog and distribute the source SNe within the source plane such that they produce multiple images. Some of these sources get doubly imaged (`doubles'), i.e., they form two images, while some get quadruply imaged (`quads'), i.e., they form four images. We use a Python pipeline from \cite{more2022improved} for generating mock lensed SNe.  We extract the lensing observables for all of the images using the software \texttt{Glafic} \citep[][]{oguri2010glafic}. The SNe light curves are simulated using the publicly available Python package, \texttt{SNCosmo} \citep[][]{barbary2016sncosmo}, using the in-built light curve templates.

\subsection{Lens galaxy population}\label{ssec:lenspop}
\begin{figure}
	\centering
	\includegraphics[width=0.85\linewidth]{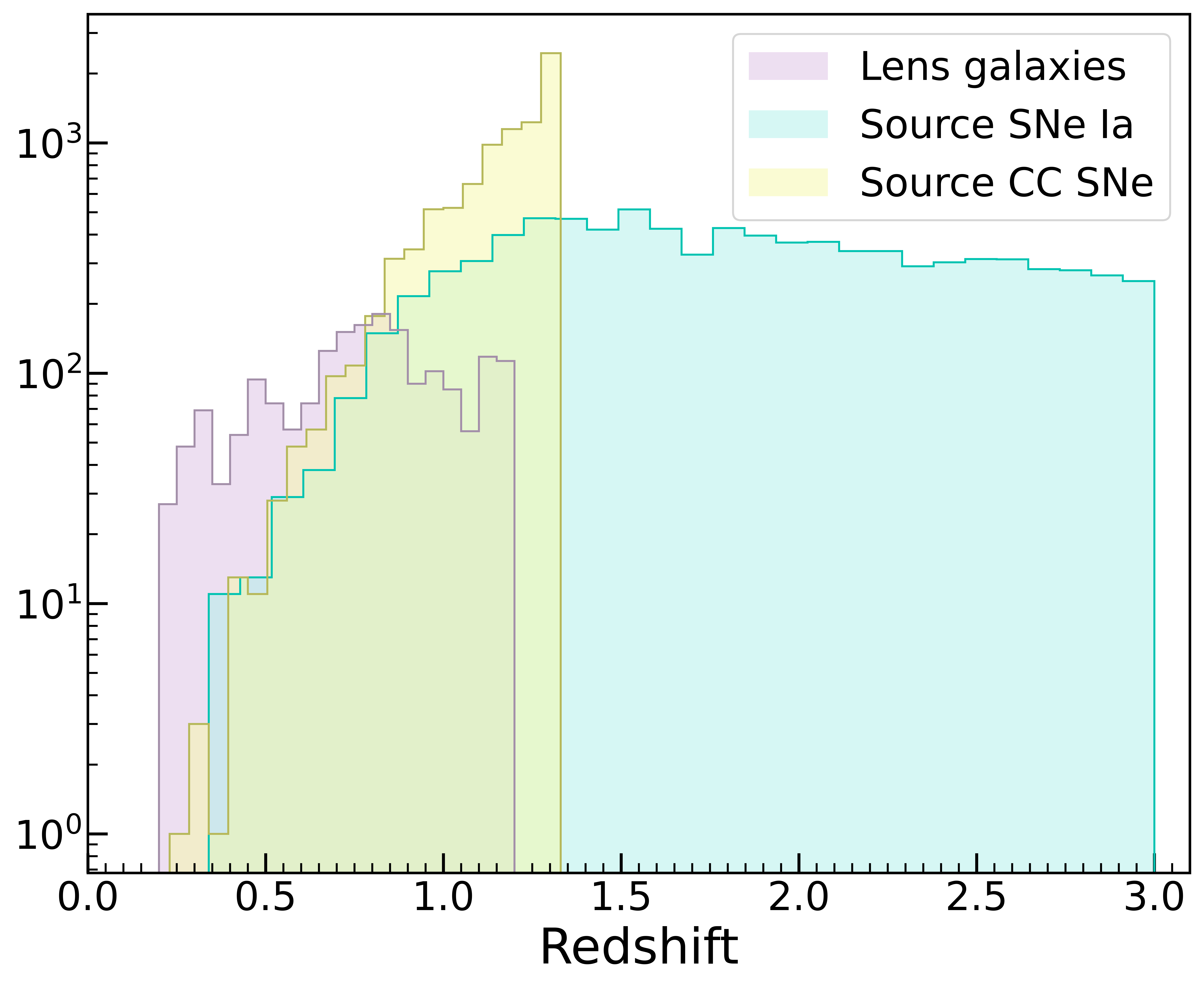}
    \caption{Distribution of the source (SNe Ia and CC SNe) and the lens redshifts used in the analysis.}
	\label{fig:z_distrib}
\end{figure}

We draw about 8,000 massive elliptical galaxies from the HSC galaxy catalog within tract 9813 as deflectors. Figure \ref{fig:z_distrib} shows the redshift distribution of this lens population. We model the density profile of lens galaxies as singular isothermal ellipsoids \citep[SIE,][]{kormann1994isothermal}. This density profile is shown to be in excellent agreement with observations \citep[e.g.,][]{koopmans2009structure}. We use observed shapes, photometric magnitudes, and redshifts of the chosen galaxies to convert their light parameters into the SIE model mass parameters. Following the $L - \sigma_v$ scaling relation \citep[table 1;][]{parker2007masses}, we obtain the velocity dispersion. Similarly, assuming mass follows light, we obtain the position of the lens potential and the ellipticity parameters. We do not consider external shear in this analysis.

Given the lens and the source redshifts, the probability that such a configuration could lead to lensing is calculated from the optical depth, given as,
\begin{equation}
    \tau = \frac{\Omega_{\rm{lens}}}{4\pi}\,,
\end{equation}
where $\Omega_{\rm{lens}}$ is the solid angle within which the source should lie to be multiply imaged. We boost this optical depth by a constant factor, an arrangement equivalent to considering each lens source pair as representative of many lens source pairs with the same distribution of properties as expected of lensed events. Using the optical depth value, we distribute the positions of the sources uniformly in the source plane such that a lensing event producing multiple images will occur. Distributing the sources uniformly ensures that the resulting sample has a realistic fraction of doubles and quads. For each lens system, we obtain the magnification factors and the time delays of the lensed images, which are further used to simulate lensed SNe light curves at the positions of the lensed images.

\subsection{Source population}
The redshift of each SN is drawn from the redshift-dependent volumetric rates, R($z$), in the local rest frame based on the fits to the observed population of SNe. For SNe Ia with redshift $<$ 1, we follow the rate distribution of \cite{dilday2008measurement}, and for SNe Ia with redshift $>$ 1, we follow the rate distribution of \cite{hounsell2018simulations}, given by,

\begin{eqnarray}
	{\rm R}(z) & = & 2.5 \times (1+z)^{1.5} \times 10^{-5} \textrm{ yr}^{-1} \textrm{ Mpc}^{-3}\,, \quad \textrm{for}\ z < 1 \\
	{\rm R}(z) & = & 9.7 \times (1+z)^{-0.5} \times 10^{-5} \textrm{ yr}^{-1} \textrm{ Mpc}^{-3}\,, \quad \textrm{for}\ z > 1\,.
\end{eqnarray}

We also simulate a set of unlensed and lensed CC SNe to study the contamination from this class of transients in the space occupied by lensed SNe Ia in the color-magnitude diagram. A CC SNe results from a sudden collapse of a star. They can be further classified into Type Ib/Ic and Type II SNe \citep[e.g.,][]{minkowski1979spectra, filippenko1997optical}. Type Ib early-time spectra show prominent He I, whereas type Ic spectra show neither Si II nor He I. Type II SNe are defined by the presence of hydrogen features in the early-time spectra. SNe II are further classified into broadly three photometric subclasses: IIL (for \textit{linear}) that generally resemble SNe I light curves, IIP (for \textit{plateau}) for which the light curve remains within $\sim$1 mag of maximum brightness for an extended period and then declines gradually, and IIn (for \textit{narrow}) that has narrow lines of hydrogen emission in spectra, implying dense pre-existing circumstellar material prior to the explosion. At late times, the light curves of most SNe II resemble each other, both in shape and luminosity. Apart from these types, a few supernovae in literature have been reported to change types: they show lines of hydrogen at early times, but over a period of weeks to months, become dominated by lines of helium. These have been designated as type IIb SNe, and they show a combination of features associated with types II and Ib. In this analysis, we study contamination by these six subclasses of CC SNe: Type Ib, Ic, IIL, IIP, IIn, and IIb. 

The redshifts of CC SNe are drawn from the redshift-dependent volumetric rates using the star formation history from \cite{madau2014cosmic}, and CC SNe rate curve is derived from figure 1 of \cite{wojtak2019magnified}. 

From R($z$), we determine the total number of observed SNe per unit time (in this case, it is per year) as a function of redshift as,
\begin{equation}
    \frac{\rm{d}N}{\rm{d}z \rm{d}t} = 4\pi \frac{c}{H_0} \frac{(1+z) D^2}{E(z)} \rm{R}(z)\,,
\end{equation}
where D is the angular diameter distance and $E(z) = H(z)/H_0$ is the dimensionless Hubble parameter. Figure \ref{fig:z_distrib} shows the redshift distribution of the SNe Ia and CC SNe population simulated in this work.

\subsection{Simulating light curves}\label{ssec:SNeIa_lc}
We use the \texttt{SALT2} light-curve model \citep[][]{guy2007salt2, guy2010supernova} to simulate SNe Ia for the entire analysis as point sources. The input parameters of the model are B-band peak amplitude ($x_0$), stretch  ($x_1$), color ($c$), redshift ($z$), and the time at B-band peak brightness ($t_0$). The stretch and color parameters, respectively, account for variations in the shape of the light curve and color as a function of luminosity (brighter-broader and brighter-bluer effects). The distance modulus of a supernova is given by 
\begin{equation}
	\mu = m_{\rm B} - M + \alpha x_1 - \beta c\,,
\end{equation}
where $m_B$ = $-2.5\log(x_0$), M is the absolute magnitude of the SN Ia with $c$ = $x_1$ = 0, and $\alpha$ and $\beta$ are nuisance parameters representing slopes of the stretch-luminosity and color-luminosity relations. 

The \texttt{SALT2} model has coverage in the spectral range of rest-frame 2000 $-$ 9200 \si{\angstrom}. The LSST bands \textit{rizy} lie between the observer-frame wavelength of 5,500 and 11,000 \si{\angstrom}. Thus, we consider sources up to redshift 3 to make sure that we can simulate photometric data in either \textit{r} and \textit{i} bands or \textit{z} and \textit{y} bands. We draw the values of the color and stretch parameter between $-$0.3 and 0.5 and $-$3 and 2, respectively, from the asymmetric distribution as described in \cite{scolnic2016measuring}, table 1 (a mean of $-$0.054 and an asymmetric scatter of $-$0.043 and $+$0.101 for $c$ and a mean of 0.973 and an asymmetric scatter of $-$1.472 and $+$0.222 for $x_1$). The amplitude parameter, $x_0$, is computed using the relation $x_0 = 10^{0.4(\alpha x_1 - \beta c-M)}$, where we fix values of the global standardization parameters $\alpha$ and $\beta$ to be 0.14 and 3.1, respectively \citep[]{betoule2014improved}. We assume that M, the peak absolute magnitude in B-band, is normally distributed with a mean of $-$19.35 and a scatter of 0.13 mag.

Unlike SNe Ia, CC SNe are far more diverse in their observable properties, and the templates for the subclasses are not as well-constrained. In this work, we simulate the light curves and K-corrections for SN Ib, Ic, IIL, IIP, and IIn subclasses of CC SNe using spectral templates from an extension of the work by \cite{nugent2002k}. For type IIb, we use the template given by \cite{vincenzi2019spectrophotometric}. All of the templates take three parameters as input: B-band peak amplitude ($x_0$), redshift ($z$), and the time at B-band peak brightness ($t_0$). The nugent templates have spectral coverage in the range of 1000 $-$ 25000 \si{\angstrom}, whereas the template for SNe IIb has coverage in the range of 1605 $-$ 11000 \si{\angstrom}. Table 2.1 lists the set of templates, luminosity functions (LFs), and relative fractions of each subclass out of the total CC SNe population we use to simulate the photometric data in this work.

\begin{table}
	\centering
        \resizebox{\columnwidth}{!}{%
	\begin{tabular}{cccc}
		\hline
		SN Type & SNCosmo Template & LF ($\bar{M_B}[\sigma]$) & Relative fraction (\%)\\
		\hline
		IIP & nugent-sn2p$^b$ & $-$16.75 [0.98] & 55.83\\
		Ic & nugent-sn1bc$^l$ & $-$17.66 [1.18] & 17.00\\
		IIb & v19-2006t-corr$^v$ & $-$16.99 [0.92] & 12.43\\
		Ib & nugent-sn1bc$^l$ & $-$17.45 [1.12] & 9.00\\
		IIL & nugent-sn2l$^g$ & $-$17.98 [0.86] & 3.34\\
		IIn & nugent-sn2n$^d$ & $-$18.53 [1.36] & 2.40\\
		\hline
	\end{tabular}%
        }
    \label{tab:CC SNe_params}
    \caption{Details of CC SNe templates, LFs, and relative fraction values used in the simulations. The LF values are reported from \protect\cite{richardson2014absolute} as the mean $\bar{M_B}$ and scatter $\sigma$ of assumed Gaussian. CC SNe relative fractions are as reported in \protect\cite{eldridge2013death}. $^b$: \protect\cite{baron2004type}; $^l$: \protect\cite{levan2005grb}; $^v$: \protect\cite{vincenzi2019spectrophotometric}; $^g$: \protect\cite{gilliland1999high}; $^d$: \protect\cite{di2002optical}.}
\end{table}

We input this information in \texttt{SNCosmo} to generate unlensed light curves of SNe Ia and CC SNe subtypes. We incorporate time delay and magnification values in the unlensed light curves manually to obtain the lensed resolved light curves. We generate the `unresolved' sample by simply considering the total flux of all lensed images of a lens system from the `resolved' sample without any criterion of angular separation between images and using the same set of lensing observables. The purpose of this is to compare samples with a consistent underlying population that only differ in the way the flux gets observed by the survey. Whether a lensed transient will get detected as resolved or unresolved depends on the observing conditions and the survey-specific properties, which we do not consider here as we focus on demonstrating the usability of color-magnitude selection on the intrinsic lensed SN population.

\section{Observed supernovae data}\label{ssec:observed_SNeIa}
We studied the simulated SNe Ia data so far with the CM diagrams (CMDs) to propose the red limit that selects the simulated lensed SNe from unlensed SNe Ia. 
To verify the efficiency of the proposed selection criterion on the observed SNe data, we use archival data from various SN Ia surveys. We use a publicly available Python package \texttt{SNData}\footnote{\href{https://sndata.readthedocs.io/en/latest/}{https://sndata.readthedocs.io/en/latest/}} to access the photometric data of 867 spectroscopically confirmed SNe Ia light curves from three SN surveys: 124 light curves from the Dark Energy Survey (DES) Data Release 1 \citep[DES-SN3YR;][]{brout2019first} up to redshift of 0.8, 211 light curves from the Equation of State: Supernovae trace Cosmic Expansion \citep[ESSENCE;][]{narayan2016light} survey up to redshift of 0.8, and 532 light curves from the Joint Light curve Analysis \citep[JLA;][]{betoule2014improved} up to redshift of 1.1. We also use 252 low redshift SNe Ia light curves from Zwicky Trasient Facility (ZTF) Data Release 2 \citep[ZTF DR2;][]{rigault2025ztf} up to redshift of 0.2, downloaded using publicly available Python package \texttt{ztfcosmo}\footnote{\href{https://github.com/ZwickyTransientFacility/ztfcosmo}{https://github.com/ZwickyTransientFacility/ztfcosmo}}. Additionally, we use light curves of 88 CC SNe accessed using \texttt{SNData} observed in the Sloan Digital Sky Survey (SDSS) \citep[SDSS-II;][]{sako2018data} with redshifts up to 0.3.

We compare this with the data of three known systems of lensed SNe for which optical photometry is available\footnote{A few more recently detected systems like SN H0pe, SN Requiem do not have publicly available optical data in \textit{r} and \textit{i}-bands}: PS1-10afx ($z$ = 1.38) with the Pan-STARRS1 telescope \citep[][]{quimby2014detection}, iPTF16geu ($z$ = 0.4) with the Palomar 60-inch telescope \citep[][]{goobar2017iptf16geu}, and SN Zwicky ($z$ = 0.35) with the Liverpool telescope \citep[][]{goobar2023uncovering}. Along with this unresolved light curve information, we also include data of four individual images from the simulated resolved light curves of iPTF16geu and SN Zwicky \citep[][]{goobar2017iptf16geu, goobar2023uncovering}.

Furthermore, we use the observed data of 42 SLSNe discovered in various surveys. SLSNe are a class of stellar explosions initially defined by their peak absolute magnitudes greater than $-$21. This corresponds to the total radiated energies of $\mathcal{O}(10^{51})$ erg, and they can be 5 to 100 times brighter than Type Ia SNe or typical core-collapse events \citep[][]{gal2012luminous}. For a detailed review of SLSNe, we refer to \cite{gal2019most}. SLSNe are a rare class of transient objects with rates about 100 times lower than ordinary supernovae. From their relatively recent discovery \citep[][]{quimby2011hydrogen}, a few hundred SLSNe have been detected in various surveys to date. LSST is expected to detect about 10$^4$ useful\footnote{\cite{villar2018superluminous} give this number estimate for SLSNe that will have at least 10 observations in LSST. We note that the estimate obtained with this stringent criterion itself is large enough to act as a contaminant for lensed SNe searches.} SLSNe per year \citep[][]{villar2018superluminous}. We compare the CM parameter space occupied by SLSNe with that of lensed SNe Ia since, due to the lensing magnification, the apparent magnitudes of lensed SNe can be comparable to those of SLSNe. This may lead to the false selection of SLSNe as lensed SNe in brightness-based lensed SNe searches. To compare the CMD of SLSNe, we used a representative sample of 42 SLSNe light curves extracted manually from various sources \citep[e.g.,][]{chomiuk2011pan, howell2013two, inserra2013super, nicholl2014superluminous, lunnan2015zooming, mccrum2015selecting, nicholl2017magnetar, inserra2018nature}.

For the observed lensed SNe Ia, we note that not only the multiply lensed images are unresolved, but also the data has sparse sampling. To tackle this, we use the data obtained by fitting a light curve template for the (un-)lensed SNe Ia sample. For the observed CC SNe and SLSNe sample, we select the observations that are closest to our chosen epochs (see Sec. \ref{ssec:epoch} for the choice of epochs).

\section{Results}\label{sec:results}
In this section, we present the CMDs for SNe Ia with varied properties and propose red limits to select lensed SNe. Our proposed criterion selects the SNe above the red limit as potential lensed SNe Ia. All the times reported in this section are in the observer frame.

\subsection{Phase of the light curves}\label{ssec:epoch}
To compare the photometric data consistently for different populations, we study the unresolved fluxes at two sets of two epochs (one epoch each on the rising and falling edge of the light curve). In Set 1, we study the data from three days before the \textit{i}-band peak - time at which the unlensed SN brightness reaches the peak in the \textit{i}-band - for the rising edge, and seven days after the \textit{i}-band peak for the falling edge. In set 2, we choose the epochs at fifteen days before and thirty-five days after the \textit{i}-band peak for the rising and falling edges, respectively. These epochs are set considering the length of a typical SN Ia light curve. In \cite{quimby2014detection}, the red limit has been determined to select lensed SNe Ia on the rising phase of the SNe light curve. However, in transient search surveys, SNe could be discovered at any point in their light curves, either in their rising or falling phases. Thus, we study the CMDs at various epochs of the light curve to understand if CM selection enables identification at any phase of the light curve.

\begin{figure}
	\centering
	\includegraphics[width=1.\columnwidth]{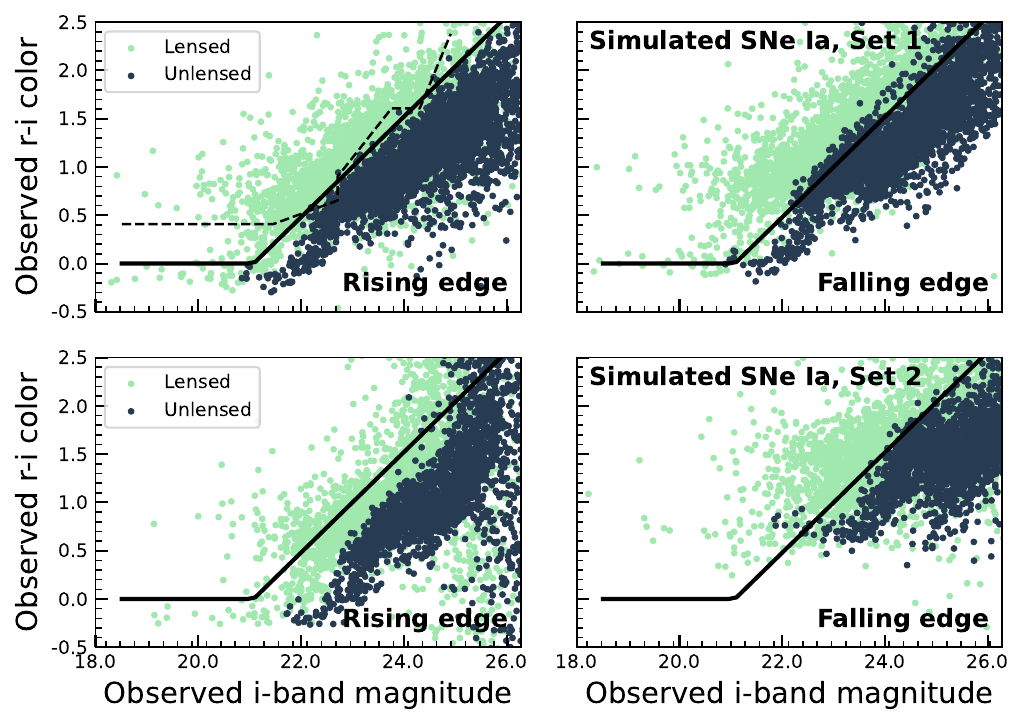}
	\caption{CMDs of unlensed and lensed SNe Ia for two sets of chosen epochs up to $z$ = 1.64 comparing the rising (left column) and falling (right column) phases of the SNe light curve for Set 1 (upper panel) and Set 2 (lower panel) for specific epochs (see the text for details). The black dashed curve is the red limit defined by \protect\cite{quimby2014detection}. The bold black curve is the proposed modified red limit for unlensed SNe Ia on the rising and falling phases of the light curve.}
	\label{fig:epoch}
\end{figure}

Figure \ref{fig:epoch} shows the CMDs of simulated SNe Ia for the two sets of epochs of observation. We note that a subset of lensed SNe Ia occupy a region in CM parameter space that does not overlap with unlensed SNe Ia, both during the early epochs (set 1) and the late epochs (set 2) of the light curves, regardless of the phase. The original red limit \citep[][]{quimby2014detection}, given by the black dashed curve, is able to separate the non-overlapping subset of lensed SNe Ia. We modify this curve slightly to better suit our Set 1 distribution and propose the modified red limit (black bold curve in figure \ref{fig:epoch}). We observe that in our sample, the proposed criterion selects lensed SNe Ia more efficiently using the modified red limit and hence continue to use this modified curve throughout the remainder of this paper.

We also study the CMDs for an additional Set 3, (thirty days before and seventy days after the peak) and observe similar trends in the CM parameter space. These results are not shown here to avoid repetition. However, we note that, a majority of unlensed SNe Ia and a significant proportion of lensed SNe Ia fall below the single-epoch depths of current and upcoming surveys for the late (Sets 2 and 3) rising and falling epochs, rendering them undetectable and thus, observations at these latter epochs will have little-to-no practical importance. Owing to this factor, we work with Set 1 epochs only in the remainder of this paper. Thus, hereafter, the \textit{rising epoch} and the \textit{falling epoch} always refer to photometric data reported three days before and seven days after the \textit{i}-band peak, respectively. 

It is important to note that the proposed black curve is an efficient selection criterion in the context of the simulations in this work. For precise selection cuts that work for a dataset with different properties, one must simulate data tailored to the required properties, such as specific survey selection criteria and observing limitations, and modify this curve accordingly. While this study does not incorporate LSST-specific properties\footnote{Except for LSST bandpasses.} such as PSF size and cadence, these factors are accounted for in our companion study (Mane et al., 2025, in prep.), where we derive a color-magnitude limit specific to LSST using the (un-)lensed SNe Ia light curves obtained upon running the difference imaging (DI) pipeline.  With this in consideration, the selection curve for Set 1 shown in figure \ref{fig:epoch} can be given as,
\begin{equation}
    \mathrm{m}_r - \mathrm{m}_i > 
    \begin{cases} 
    0.52 \, \mathrm{m}_i - 10.96 & \text{if } \mathrm{m}_i > 21.02 \\
    0.0 & \text{otherwise}
    \end{cases}
\end{equation}

We find that this curve selects $\approx$ 44\% of lensed SNe Ia while rejecting $\approx$ 99\% of unlensed SNe Ia on the rising edge, and selects $\approx$ 67\% of lensed SNe Ia while rejecting $\approx$ 91\% of unlensed SNe Ia on the falling edge. 

We note that modifying the proposed curve slightly on both rising and falling edges results in the selection of slightly higher lensed SNe and slightly lower unlensed contaminants, respectively. However, this improvement is small, and the effectiveness of the proposed curve in selecting lensed SNe Ia on their falling edge is also promising. This is particularly encouraging given that supernovae are more likely to be detected at their peak with follow-up observations during their falling phases. The fact that CM-based selection remains effective in this phase ensures that the method remains useful even when early-time rising phase observations are not available. Additionally, the criteria remaining useful at early epochs equally well will allow timely identification and follow-up triggers if a lensed SNe Ia is detected at rising epochs.

\subsection{Source Redshift}
For each unlensed parent supernova and its images, we calculate the apparent magnitudes and colors using the \texttt{SALT2} model. The model offers observer-frame flux coverage in the LSST \textit{r} and \textit{i}-bands for SNe with $z<1.64$. We use the \textit{i}-band magnitude and \textit{r}$-$\textit{i} color for SNe Ia with $z<1.64$ (\textit{low} redshift bin) and consider the use of the \textit{y}-band magnitude and \textit{z}$-$\textit{y} color for those with redshifts $1.64<z<3$ (\textit{high} redshift bin). This allows us to set the red limit for SNe Ia at higher redshifts that may be detectable in LSST. 
 
\begin{figure}
	\centering
	\includegraphics[width=0.9\linewidth]{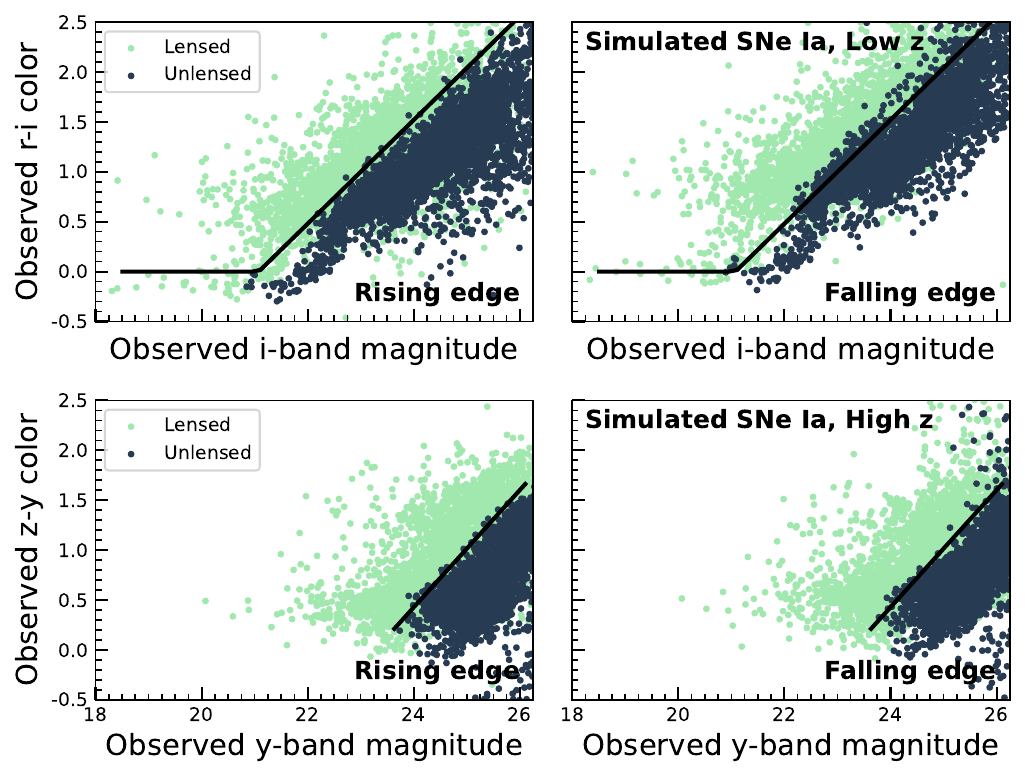}
	\caption{CMDs of unlensed and lensed SNe Ia comparing the red limits for SNe Ia at low (upper panel; $z<1.64$) and high (lower panel; $1.64<z<3$) redshifts. We propose the use of observer-frame {\textit y}-band magnitude and {\textit z}-{\textit y} color to draw CMDs for SNe Ia at high redshifts. The left and right panels compare the rising and falling phases of SNe Ia light curves. The bold black curves in both panels give the red limit for unlensed SNe Ia until $z<3$.}
	\label{fig:redshift}
\end{figure}

Figure \ref{fig:redshift} shows the CMDs for the simulated SNe Ia at low and high redshifts. We note that, similar to the case of low redshift SNe Ia in the \textit{r}$-$\textit{i} versus \textit{i} panel, a subset of even high redshift lensed SNe Ia occupy a non-overlapping region of CM parameter space with the high redshift unlensed SNe Ia in the \textit{z}$-$\textit{y} versus \textit{y} panel. This allows us to determine the red limit for the high-redshift SNe Ia sample as well. Based on this observation, we propose the use of the \textit{z} and the \textit{y} bands to identify higher redshift lensed SNe Ia using the black bold curve in figure \ref{fig:redshift}. 

The selection curve for high-z sample shown in the figure can be given as,
\begin{equation}
    \mathrm{m}_z - \mathrm{m}_y > 
    \begin{cases} 
    0.59 \, \mathrm{m}_y - 13.67 & \text{if } \mathrm{m}_y > 23.63 \\
    0.0 & \text{otherwise}
    \end{cases}
\end{equation}
This curve selects $\approx$ 46\% of lensed SNe Ia at high redshift while rejecting $\approx$ 99.5\% of unlensed SNe Ia on the rising edge, and selects $\approx$ 45\% of lensed SNe Ia while rejecting $\approx$ 98\% of unlensed SNe Ia on the falling edge.

\subsection{Unresolved versus resolved}
The red limit proposed by \cite{quimby2014detection} was originally meant for unresolved lensed SNe Ia. We extend this further to include resolved lensed SNe and test whether this red limit will hold. Figure \ref{fig:resolved} shows the CMDs for simulated resolved SNe Ia at low and high redshifts and for both the rising and falling edges. Comparing this with figure \ref{fig:redshift}, we note that the resolved SNe Ia are also selected successfully using the same red limit that selects unresolved SNe Ia. As expected, the resolved lensed SNe Ia distribution shifts to fainter \textit{i}-band magnitudes than the unresolved lensed SNe population. This is because the unresolved lensed SNe magnitudes are the combined magnitudes of the multiple lensed images of SNe, making them typically brighter than the magnitudes of the individual lensed images or any unlensed SNe at the same redshift. We note that the selection criterion can more robustly identify unresolved lensed SNe Ia. For the resolved lensed SNe Ia, while this criterion is equally successful in eliminating unlensed SNe Ia, it misses a higher fraction of lensed SNe Ia. Following this result, and because the unresolved systems are expected to dominate the detections \citep{quimby2014detection} in a ground-based survey, we work with only unresolved SNe light curves in the rest of the paper.

\subsection{Contamination from core-collapse supernovae}
We study simulated unlensed and lensed CC SNe to understand their distribution in the CMDs and assess their potential to contaminate lensed SNe Ia sample under our selection criteria. Studying this contamination is necessary given that the type of a detected supernova will often be unknown. For this, we simulate CC SNe data with realistic assumptions as described in Sec. \ref{sec:SNe_sim}. 
  
\begin{figure}
	\centering
	\includegraphics[width=1.\linewidth]{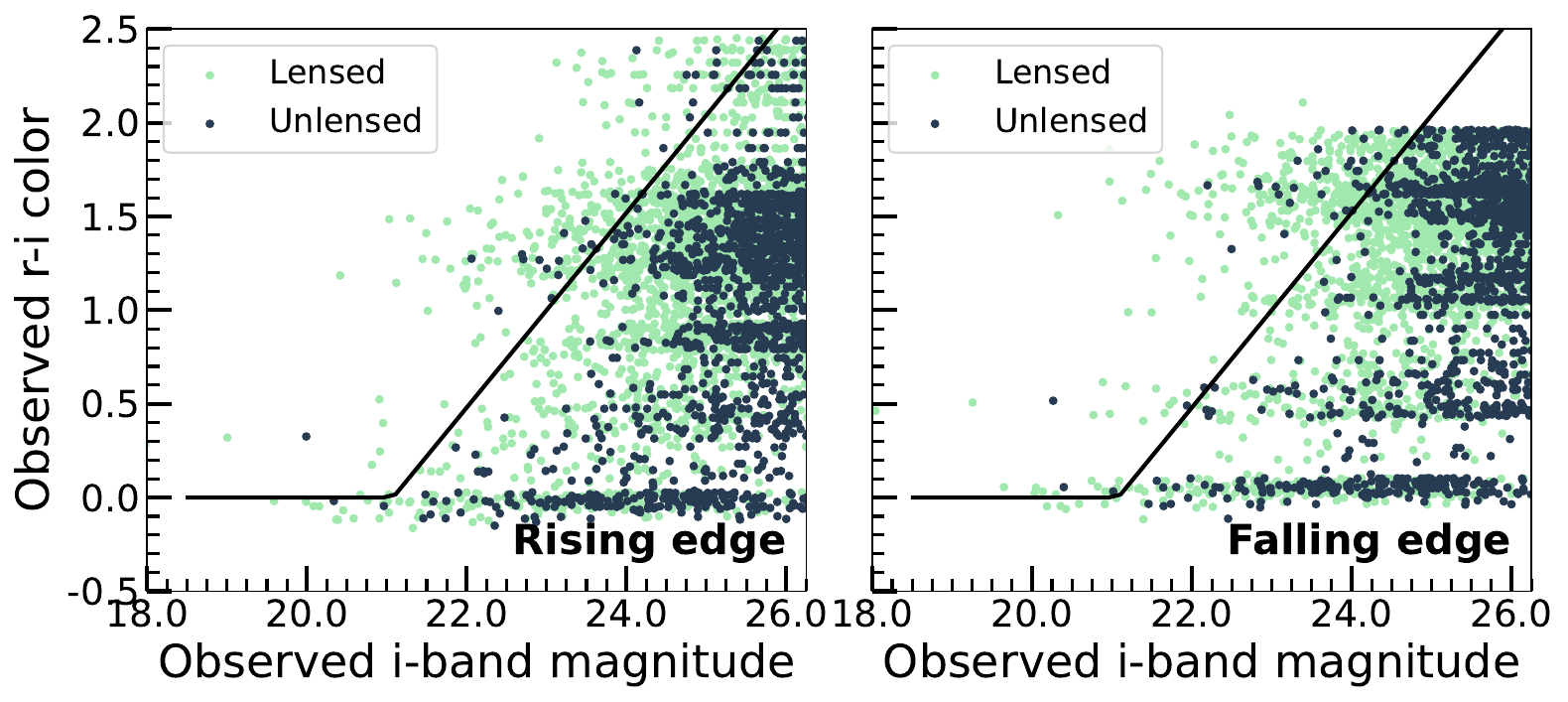}
	\caption{CMDs of the simulated lensed and unlensed CC SNe for $z<1.33$ comparing both the rising (left panel) and falling (right panel) edges. The red limit (black bold curve) that separates lensed SNe Ia from the unlensed also eliminates almost all of the unlensed CC SNe, but may allow a small fraction of lensed CC SNe with fainter magnitudes to be selected.}
	\label{fig:sncc_together}
\end{figure}

Figure \ref{fig:sncc_together} shows the CMD for unlensed and lensed CC SNe (unresolved) for $z<1.3$. We note that the contamination by unlensed CC SNe is very low to negligible ($<$ 0.05\%), while the contribution from lensed CC SNe is slightly more ($\approx$ 1\%). Figure \ref{fig:sncc_separate} shows the CMD for only lensed CC SNe, with the different CC SNe types color-coded differently. Here, we see that lensed SNe Ib and Ic from among CC SNe fall above the red limit.

We compare the low redshift (for $z<1.33$) CC SNe data in this analysis, as most of the CC SNe types are typically fainter than SNe Ia, and hence, we expect to detect those types of SNe primarily from nearby host galaxies. Apart from observational limitations, the LFs and fractions of different types of CC SNe, and the spectral templates of individual types are less certain and may be a source of additional systematic errors \citep[please see ][for various observational constraints on these parameters]{smartt2009death, eldridge2013death, richardson2014absolute, shivvers2017revisiting}, especially at higher redshifts. With consideration of these limitations, we see that the CM criteria can potentially select a small fraction of lensed CC SNe -- SNe Ib and Ic are more likely -- while expecting negligible contamination from unlensed CC SNe.

\subsection{Comparison with the observed supernovae data}
\begin{figure}
	\centering
	\includegraphics[width=1.0\linewidth]{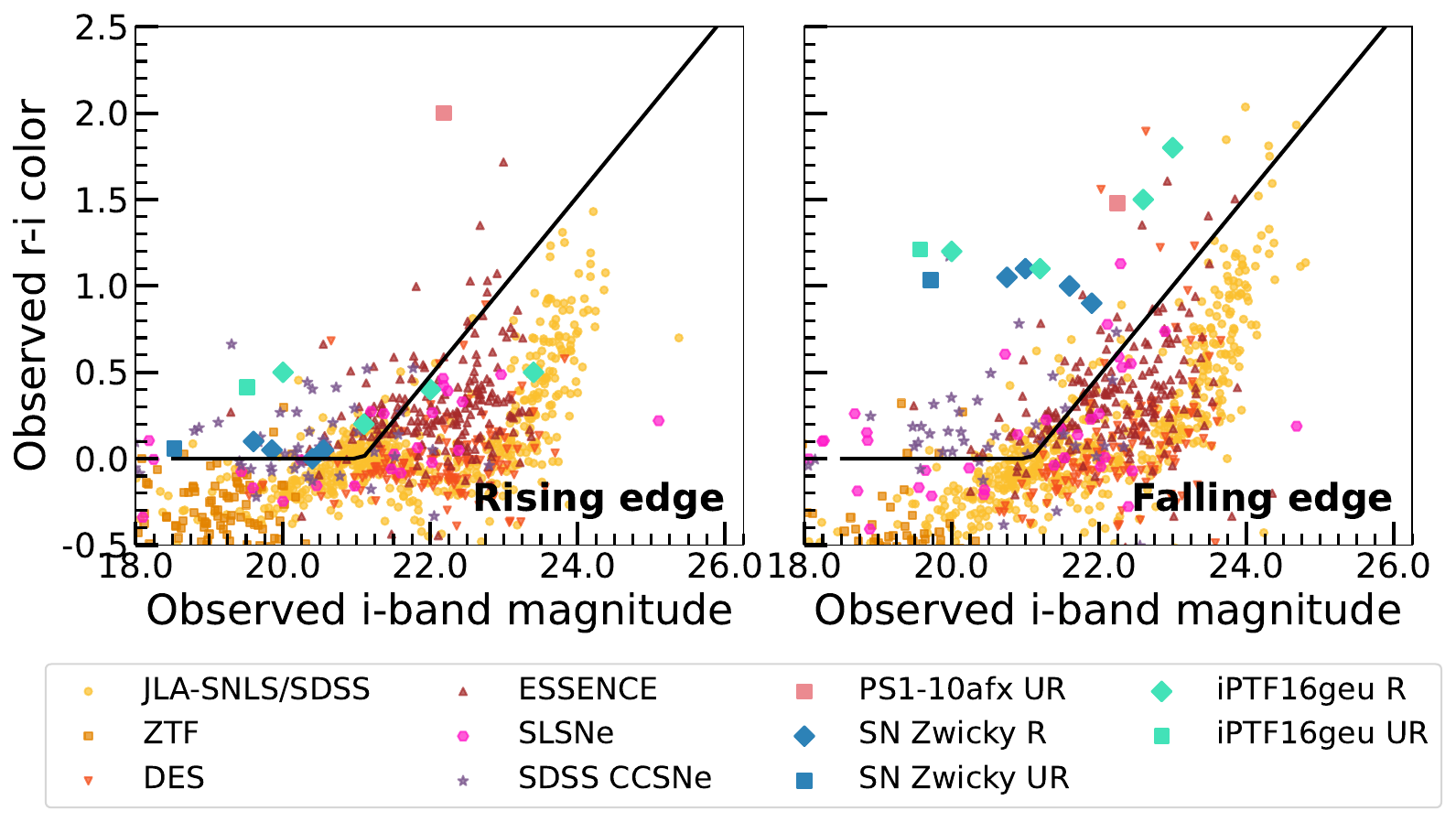}
	\caption{CMDs of the observed unlensed SNe Ia, CC SNe, and SLSNe along with the small sample observed lensed SNe to verify the validity of the proposed criterion for real SNe. The proposed selection criterion eliminates a majority of unlensed SNe Ia, and SLSNe while selecting the known lensed SNe Ia systems (both the unresolved and resolved light curves) on the falling edges of the respective light curve.}
	\label{fig:observedSNIa}
\end{figure}

To verify the applicability of the proposed selection criterion on the observed SNe data, we find a variety of SNe Ia, CC SNe, and SLSNe from surveys in the literature, as detailed in section \ref{ssec:observed_SNeIa}. The CMDs of these SNe are shown in figure \ref{fig:observedSNIa}. 

We note that the photometry of each of the SNe in this figure is recorded in the \textit{i} and \textit{r} bandpass filters of the respective survey telescopes, which may differ from the LSST bandpasses for which the selection cut in consideration is originally devised. Thus, it is important to consider the further modification that the black curve in figure \ref{fig:epoch} will undergo for the respective telescope bandpasses, the effect that we do not treat explicitly here but affects the suitability of the proposed selection criterion when testing it on data from surveys other than LSST. With this in consideration, we note from figure \ref{fig:observedSNIa} that our selection criterion eliminates a majority of unlensed SNe Ia and the SLSNe while selecting the known lensed SNe Ia systems (both unresolved and resolved light curves) on the falling edges of respective light curves. The cut is, however, found to select a majority of the low redshift unlensed CC SNe sample. We note that we have tested a very small sample of CC SNe from a single survey in this work. A higher, more realistic fraction of CC SNe at lower redshift with well-observed light curves will be necessary to study if real CC SNe indeed contaminate lensed SNe Ia selected by the CM criterion.

\section{Summary and conclusions}\label{sec:conclusion}
With the onset of many wide and deep transient imaging surveys such as the Rubin LSST, it is crucial to have fast lens search algorithms or tools that can rapidly reject most of the unlensed transients and help identify potential lensed SNe Ia. This will enable us to quickly trigger suitable follow-up spectroscopy or deploy a more accurate lens search pipeline that may not be fast enough to process $\mathcal{O}(10^6)$ daily alerts \citep[][]{abell2009lsst}. With this goal, we revisited and investigated the suitability of a simple color-magnitude (CM) criterion to select promising lensed SNe Ia candidates from among other unlensed SNe (and ultimately, all transients) that the alert pipelines may report. 

\cite{quimby2014detection} originally proposed the use of CM selection criterion to distinguish lensed SNe Ia from unlensed SNe. They proposed a red limit in the color-magnitude diagram (CMD) that selects unresolved lensed SNe candidates on the rising edge of their light curve. In this study, we further extended this analysis and proposed modifications to the red limit for SNe datasets under different conditions. We demonstrated that the red limit can help identify promising lensed SNe Ia candidates for lensed sources out to $z=3$, extending the results of \cite{quimby2014detection} which had sources up to $z=2$. We showed that this criterion effectively selects both resolved and unresolved lensed SNe Ia regardless of the epoch of the light curve at which the selection criterion is applied. 

Depending upon the redshift of the source SNe Ia, different combinations of bands for the color and magnitude are suitable to identify lensed SNe Ia. In this work, we proposed the use of \textit{z} and \textit{y} band photometry for higher redshift SNe Ia CMDs. We demonstrated that this choice effectively selects lensed SNe Ia from among the unlensed population at high source redshifts ($z>2.4$). We also studied the contamination by the (un-)lensed CC SNe in the CMDs of lensed SNe Ia. We found the contamination by the unlensed CC SNe to be negligible, whereas the criterion was observed to select a fraction of lensed CC SNe, primarily of the types Ib and Ic, as potential lensed SNe Ia. This indicates that when applied to the real data, the criterion can potentially detect some lensed CC SNe systems along with lensed SNe Ia. Finally, we verified that the proposed criterion is applicable to the observed SNe using the published sample of real unlensed and lensed SNe Ia and a small sample of CC SNe and SLSNe observations.

Microlensing by the stellar population embedded within the lensing galaxy can cause further deviations in the (de-)magnifications of the strongly lensed SNe images and hence, affect their magnitudes \citep[e.g.,][]{dobler2006microlensing}. \cite{goldstein2018precise} and \cite{huber2021holismokes} further show that the effect of microlensing on strongly lensed SNe Ia is achromatic until shortly after the peak brightness, making the early-time lensed SNe Ia colors less affected by microlensing. Our goal, in this work, was to first establish that the red limit is a reasonable selection criterion for strongly lensed SNe Ia without additional complexities such as microlensing. Not to mention, determining microlensing effects accurately is computationally challenging and requires making several assumptions or model parameter choices that are likely uncertain \citep[e.g.,][]{foxley2018impact, huber2021holismokes}. As a result, we defer the implementation of microlensing to a future study. Nevertheless, we note that \cite{arendse2024detecting} explore a similar CM-based selection criterion for lower redshift ($z < 1.64$) SNe Ia with microlensing effects, and their red limit is comparable to the one proposed in this work.

Lastly, we focused here on demonstrating the suitability of the red limit in identifying lensed SNe Ia population using LSST filters. In a companion study, we are working with real Hyper Suprime-Cam (HSC) Survey data wherein simulated lensed and unlensed SNe with the LSST cadence strategy are injected. We then run the Rubin LSST Difference Imaging (DI) pipeline on the multi-epoch, multi-band HSC data to recover the simulated lensed transients as well as real artifacts that act as false positives and study them. In the near future, we also plan to inject these simulated SNe in the Rubin LSST-like DP0.2 images, recover the SNe with the DI pipeline and produce a CMD by quantifying their detected color and magnitudes. This will allow us to test the influence of the measurement uncertainties on the CM selection cuts. In this follow-up work, we intend to study realistic proportions of lens and unlens samples to better quantify the realistic efficiency of the CM-based criterion, given the high rate of alerts expected in LSST. Eventually, we would like to extend the unlensed population to include other kinds of transients, such as tidal disruption events, variable stars, or quasars, to understand the dominant source of contaminants in this approach.

\section*{Acknowledgments}
We thank Ayan Mitra, Robert Quimby, G. C. Anupama, Nikki Arendse, and Jasjeet Singh Bagla for helpful discussions and comments. We thank the referee for their constructive comments. We acknowledge the use of the high-performance computing facility Pegasus at IUCAA to carry out all the simulations. PM acknowledges the support from the INSPIRE-SHE scholarship funded by the Department of Science and Technology (DST), Government of India. AM acknowledges the support from the SERB Power Grant (SPG/2022/001866) funded by the DST.

\section*{Data Availability}
All simulated data are available from the corresponding author upon request.

\bibliographystyle{aa}
\bibliography{citations}

\appendix
\renewcommand{\thefigure}{A\arabic{figure}}
\setcounter{figure}{0}

\section{Additional CMDs}\label{sec:appendix}
Additional CMDs showing the comparison between resolved and unresolved SNe Ia and CC SNe separated by their subtypes are added here instead of showing them in the main text.

\begin{figure*}
    \centering
    \includegraphics[width=0.8\linewidth]{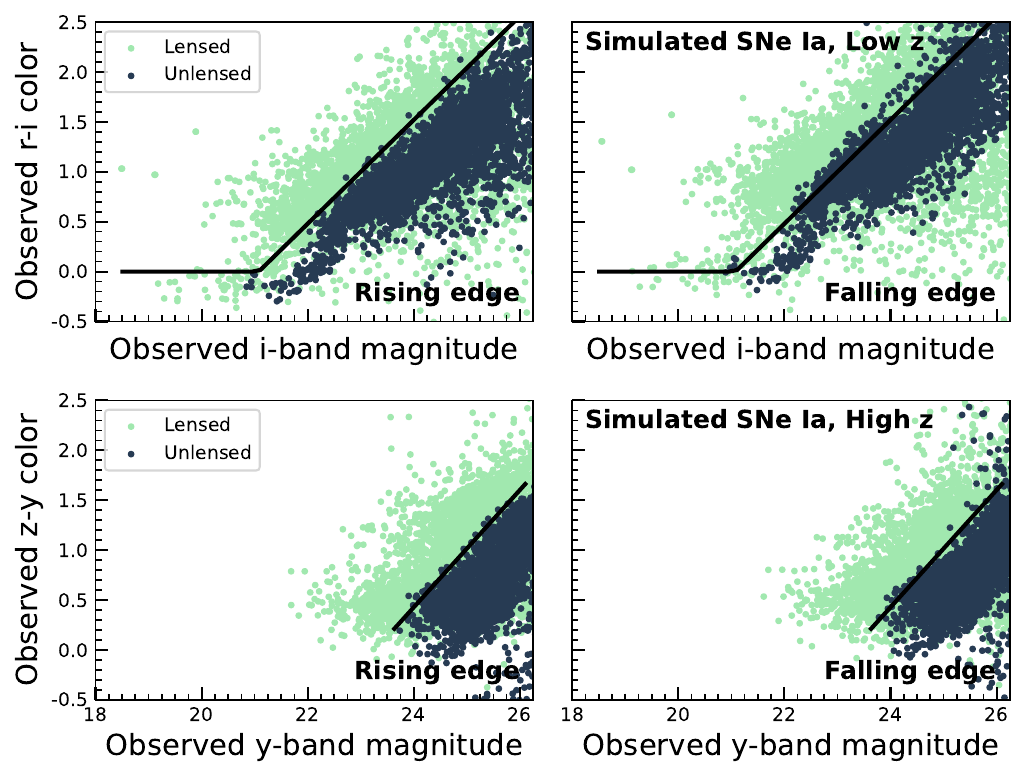}
    \caption{CMDs of unlensed and resolved lensed SNe Ia. The increased number of lensed data points compared to that in figure \ref{fig:redshift} results from considering the individual multiple images of the resolved SNe systems. The resolved lensed SNe Ia distribution is also observed to lie on a slightly fainter side compared to the unresolved SNe Ia. The red limit (bold black curve) separates resolved lensed SNe Ia from unlensed SNe Ia.}
    \label{fig:resolved}
\end{figure*}

\begin{figure*}
    \centering
    \includegraphics[width=0.8\linewidth]{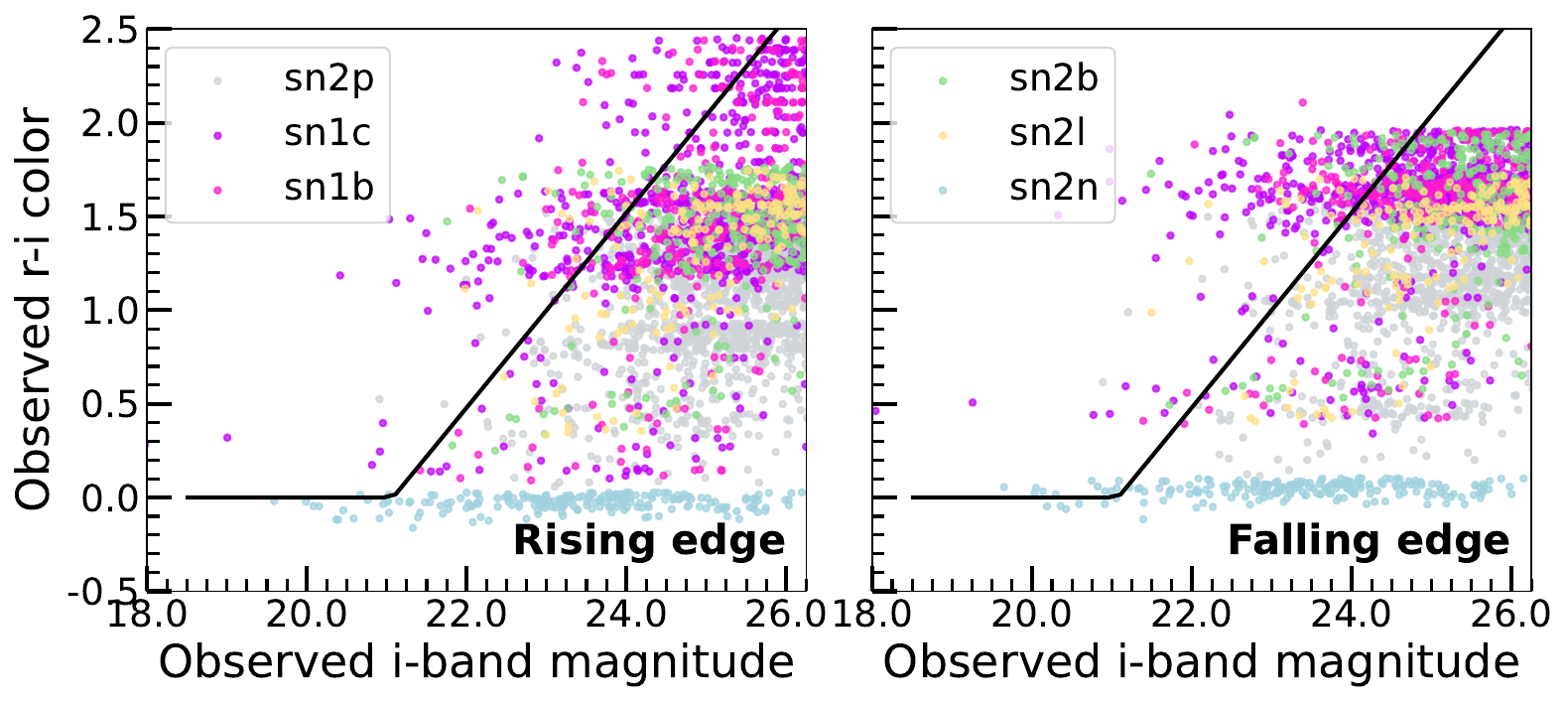}
    \caption{CMDs of the simulated lensed CC SNe for $z<1.33$ separated by six individual types. The lensed CC SNe of type Ib and Ic fall above the red limit (black bold curve), making it possible to detect at least a fraction of them with our criterion.}
    \label{fig:sncc_separate}
\end{figure*}

\end{document}